\documentclass[nocopyrightspace]{sigplanconf}

\usepackage{SIunits}            
\usepackage{courier}            
\usepackage[scaled]{helvet} 
\usepackage{url}                  
\usepackage{listings}          
\usepackage{enumitem}      
\usepackage[colorlinks=true,allcolors=blue,breaklinks,draft=false]{hyperref}   
\newcommand{\doi}[1]{doi:~\href{http://dx.doi.org/#1}{\Hurl{#1}}}   

\usepackage{graphicx}
\usepackage{xspace}
\usepackage{color}
\usepackage[table]{xcolor}
\usepackage{subcaption}
\usepackage[normalem]{ulem}

\definecolor{lightest-gray}{gray}{0.95}
\definecolor{light-gray}{gray}{0.8}

\newcommand{\code}[1]{\textsf{\small#1}}
\newcommand{\bcode}[1]{\texttt{#1}}
\newcommand{\rfmt}[1]{\textsc{#1}}

\newcommand{\sketch}{\rfmt{Sketch}\xspace}
\newcommand{\name}{\rfmt{JSketch}\xspace}
\newcommand{\pasket}{\rfmt{Pasket}\xspace}

\begin{document}

\conferenceinfo{ESEC/FSE '15}{Aug 31--Sep 4, Bergamo, Italy}
\copyrightyear{2015}
\copyrightdata{[to be supplied]}

\title{
\name{}: Sketching for Java
}

\authorinfo
{Jinseong Jeon$^\dagger$ \and Xiaokang Qiu$^\ddagger$ \and Jeffrey S. Foster$^\dagger$ \and Armando Solar-Lezama$^\ddagger$}
{$^\dagger$University of Maryland, College Park \hspace*{3em} $^\ddagger$MIT CSAIL}
{\{jsjeon@cs.umd.edu, xkqiu@csail.mit.edu, jfoster@cs.umd.edu, asolar@csail.mit.edu\}}


\maketitle

\begin{abstract}
%
Sketch-based synthesis, epitomized by the \sketch{} tool, lets
developers synthesize software starting from a \emph{partial program}, also
called a \emph{sketch} or \emph{template}.
This paper presents \name{}, a tool that brings sketch-based synthesis
to Java. \name{}'s input is a partial Java program that may include
\emph{holes}, which are unknown constants, \emph{expression
  generators}, which range over sets of expressions, and \emph{class
  generators}, which are partial classes. \name{} then translates the
synthesis problem into a \sketch{} problem; this translation is complex
because \sketch{} is not object-oriented. Finally, \name{} synthesizes
an executable Java program by interpreting the output of \sketch{}.
\end{abstract}




\section{Introduction}

\emph{Program synthesis}~\cite{syn, ded-syn} is an attractive programming
paradigm that aims to automate the development of complex pieces of code.
Deriving programs completely from scratch given only a declarative
specification is very challenging for all but the simplest algorithms,
but recent work has shown that the problem can be made tractable by starting
from a partial program---referred to in the literature as a sketch~\cite{sketch},
scaffold~\cite{pts} or template---that constrains the space of possible
programs the synthesizer needs to consider. This approach to synthesis
has proven useful in a variety of domains including
program inversion~\cite{inv},
program deobfuscation~\cite{oracle},
development of concurrent data-structures~\cite{concurrent}
and even automated tutoring~\cite{autograder}.

This paper presents \name{}, a tool that makes sketch-based synthesis
directly available to Java programmers. \name{} is built as a frontend
on top of the \sketch{} synthesis system, a mature synthesis tool based on
a simple imperative language that can generate C code~\cite{sketch}. \name{} allows
Java programmers to use many of the \sketch{}'s synthesis features,
such as the ability to write code with unknown constants (\emph{holes}
written \code{??}) and unknown expressions described by
a \emph{generator} (written \code{$\{| \ e^{*} \ |\}$}).
In addition, \name{} provides a new synthesis feature---a class-level
generator---that is specifically tailored for object oriented programs.
Section~\ref{sec:overview} walks through \name{}'s input and output,
along with a running example.

\begin{figure}[t!]
  \centering
    \includegraphics[width=\columnwidth]{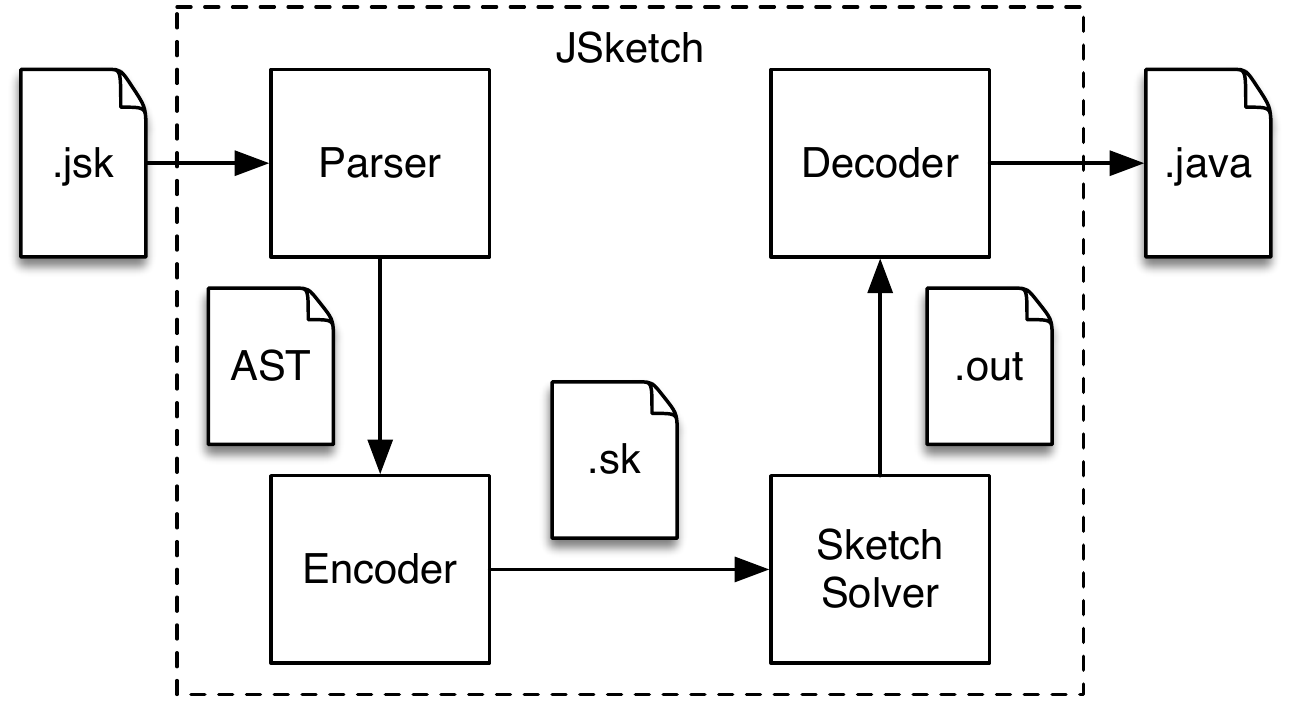}
  \caption{\name{} Overview.}
  \label{fig:overview}
\end{figure}

As illustrated in Figure~\ref{fig:overview},
\name{} compiles a Java program with unknowns to a partial
program in the \sketch{} language and then maps the result of
\sketch{} synthesis back to Java.
The translation to \sketch{} is challenging because \sketch{} is not object
oriented, so the translator must model the complex object-oriented features
in Java---such as inheritance, method overloading and overriding,
anonymous/inner classes---in terms of the features available in \sketch{}.
Section~\ref{sec:java-sketch} briefly explains several technical challenges addressed
in \name{}.
Section~\ref{sec:experience} describes our experience with \name{}.
\name{} is available at \url{http://plum-umd.github.io/java-sketch/}.

\section{Overview}
\label{sec:overview}

We begin our presentation with two
examples showing \name{}'s key features and usage.

\subsection{Basics}
\label{sec:simplemath}

The input to \name{} is an ordinary Java program that may also contain
unknowns to be synthesized. There are two kinds of unknowns:
\emph{holes}, written \code{??}, represent unknown integers and booleans, and
\emph{generators}, written \code{$\{| \ e^{*} \ |\}$}, range over a
list of expressions. For example, consider the following Java
sketch\footnote{\url{https://github.com/plum-umd/java-sketch/blob/master/test/benchmarks/t109-mult2.java}},
similar to an example from the \sketch{} manual~\cite{sketch:manual}:
\begin{lstlisting}[name=Ex]
class SimpleMath {
    static int mult2(int x) { return (?? * {| x , 0 |}); }
}
\end{lstlisting}
Here we have provided a template for the implementation of method
\code{mult2}: The method returns the product of a hole and either parameter
\code{x} or 0. Notice that even this very simple sketch has $2^{33}$ possible
instantiations (32 bits of the hole and one bit for the choice of
\code{x} or \code{0}). 

To specify the solution we would like to synthesize, we
provide a \emph{harness} containing assertions about the \code{mult2}
method:
\begin{lstlisting}[name=Ex]
class Test {
    harness static void test() { assert(SimpleMath.mult2(3) == 6); }
}
\end{lstlisting}

Now we can run \name{} on the sketch and harness.
\begin{small}
\begin{verbatim}
$ ./jsk.sh SimpleMath.java Test.java
\end{verbatim}
\end{small}
\noindent
The result is a valid Java source file in which holes and generators
have been replaced with the appropriate code.
\begin{small}
\begin{verbatim}
$ cat result/java/SimpleMath.java
class SimpleMath { ...
static public int mult2 (int x) {
return 2 * x;
}
}
\end{verbatim}
\end{small}


\subsection{Finite Automata}

Now consider a harder problem: suppose we want to synthesize a finite
automaton given sample accepting and rejecting inputs.\footnote{Of
  course, there are many better ways to construct finite
  automata---this example is only for expository purposes.}
 There are many
possible design choices for finite automata in an object-oriented
language, and we will opt for one of the more efficient ones:
the current automaton state will simply be an integer, and a
series of conditionals will encode the transition function.

Figure~\ref{fig:automaton-sketch} shows our automaton sketch. The
input to the automaton will be a sequence of \code{Token}s, which have
a \code{getId} method returning an integer (line~\ref{line:generator}). An \code{Automaton} is a
class---ignore the \code{generator} keyword for the moment---with
fields for the current \code{state} (line~\ref{line:init-st})
and the number of states (line~\ref{line:num-st}).
Notice these fields are initialized to holes, and thus the
automaton can start from any arbitrary state
and have an arbitrary yet minimal number of states
(restricted by \sketch{}'s \code{minimize} function
on line~\ref{line:min-num-st}).
The class includes a \code{transition} function that asserts that the
current state is in-bounds~(line~\ref{line:check-state})
and updates \code{state} according to the
current state and the input \code{Token}'s value~(retrieved on line~\ref{line:getid}).

Here we face a challenge, however: we do not know the number of
automaton states or tokens, so we have no bound on the number of
transitions. To solve this problem, we use a feature that \name{} inherits from \sketch{}:
the term \code{minrepeat \{ e \}} expands to the minimum length
sequence of \code{e}'s that satisfy the harness. In this case, the
body of \code{minrepeat} (line~\ref{line:st-transition}) is a
conditional that encodes an arbitrary transition---if the guard
matches the current state and input token, then the state is updated
and the method returns. Thus, the \code{transition} method will be
synthesized to include however many transitions are necessary.

Finally, the \code{Automaton} class has methods \code{transitions}
and \code{accept}; the first performs multiple transitions based on a sequence of input
tokens, and the second one determines whether the automaton is
in an accepting state. Notice that the inequality
(line~\ref{line:accept}) means that states 0 up to some bound will be
accepting; this is fully general because the exact state numbering
does not matter, so the synthesizer can choose the accepting states to
follow this pattern.


\begin{figure}
\small

  \begin{subfigure}{\columnwidth}

\begin{lstlisting}[name=Ex]
interface Token{ public int getId(); }
generator class Automaton { /** \label{line:generator} */
    private int state = ??; /** \label{line:init-st} */
    static int num_state = ??; /** \label{line:num-st} */
    harness static void min_num_state() { minimize(num_state); } /** \label{line:min-num-st} */
    public void transition(Token t) { /** \label{line:transition} */
        assert 0 <= state && state < num_state; /** \label{line:check-state} */
        int id = t.getId(); /** \label{line:getid} */
        minrepeat { /** \label{line:minrepeat} */
            if (state == ?? && id == ??) { state = ??; return;/** \label{line:st-transition} */}
    }   }
    public void transitions(Iterator<Token> it) { /** \label{line:transitions} */
        while (it.hasNext()) { transition(it.next()); }
    }
    public boolean accept() { return state <= ??; } /** \label{line:accept} */
}
\end{lstlisting}

  \caption{Automaton sketch.}
  \label{fig:automaton-sketch}
  \end{subfigure}

  \smallskip{}

  \begin{subfigure}{\columnwidth}

\begin{lstlisting}[name=Ex]
class DBConnection {
    class Monitor extends Automaton { /** \label{line:generator2} */
        final static Token OPEN =
            new Token() { public int getId() { return 1; } };
        final static Token CLOSE =
            new Token() { public int getId() { return 2; } };
        public Monitor() { }
    }
    Monitor m;
    public DBConnection() { m = new Monitor(); }
    public boolean isErroneous() { return ! m.accept(); }
    public void open() { m.transition(Monitor.OPEN); } /** \label{line:call-transition1} */
    public void close() { m.transition(Monitor.CLOSE); } /** \label{line:call-transition2} */
}
class CADsR extends Automaton { ... /** \label{line:generator1} */
    public boolean accept(String str) { /** \label{line:accept-str} */
        state = init_state_backup;
        transitions(convertToIterator(str)); /** \label{line:to-iter} */
        return accept(); /** \label{line:call-accept} */
}   }
\end{lstlisting}

  \caption{Code using Automaton sketch.}
  \label{fig:automata}
  \end{subfigure}

  \caption{Finite automata with \name{}.}
  \label{fig:template}

\end{figure}

\begin{figure}

\begin{lstlisting}[name=Ex]
class TestDBConnection {
    harness static void scenario_good() { /** \label{line:harness-good} */
        DBConnection conn = new DBConnection();
        assert ! conn.isErrorneous();
        conn.open();  assert ! conn.isErroneous();
        conn.close(); assert ! conn.isErroneous(); }
    // bad: opening more than once
    harness static void scenario_bad1() { /** \label{line:harness-bad1} */
        DBConnection conn = new DBConnection();
        conn.open(); conn.open(); assert conn.isErroneous(); }
    // bad: closing more than once
    harness static void scenario_bad2() { /** \label{line:harness-bad2} */
        DBConnection conn = new DBConnection();
        conn.open();
        conn.close(); conn.close(); assert conn.isErroneous();
}   }
class TestCADsR {
    // Lisp-style identifier: c(a|d)+r
    harness static void examples() { /** \label{line:harness1} */
        CADsR a = new CADsR();
        assert ! a.accept("c");  assert ! a.accept("cr");
        assert a.accept("car");  assert a.accept("cdr");
        assert a.accept("caar"); assert a.accept("cadr");
        assert a.accept("cdar"); assert a.accept("cddr");
}   }
\end{lstlisting}

\caption{Automata use cases.}
\label{fig:sample}
\end{figure}

\paragraph*{Class Generators.}

In addition to basic \sketch{} generators like we saw in the
\code{mult2} example, \name{} also supports \emph{class generators},
which allow the same class to be instantiated differently in different
superclass contexts. In Figure~\ref{fig:automaton-sketch}, the
\code{generator} annotation on line~\ref{line:generator} indicates
that \code{Automaton} is such a class. (Class generators are analogous to the
the function generators introduced by \sketch{}~\cite{sketch:manual}.)

Figure~\ref{fig:automata} shows two classes that inherit from
\code{Automaton}. The first class,
\code{DBConnection},
has an inner class \code{Monitor} that inherits from
\code{Automaton}. The \code{Monitor} class defines two tokens,
\code{OPEN} and \code{CLOSE}, whose ids are 1 and 2,
respectively.  The outer class has a \code{Monitor} instance
\code{m} that transitions when the database is opened
(line~\ref{line:call-transition1}) and when the database is
closed (line~\ref{line:call-transition2}). The goal is to synthesize
\code{m} such that it acts as an inline reference monitor to check
that the database is never opened or closed twice in a row, and is
only closed after it is opened. The harnesses in \code{TestDBConnection}
in Figure~\ref{fig:sample} describe both good and bad behaviors.

The second class in Figure~\ref{fig:automata},
\code{CADsR}, adds a new
(overloaded) \code{accept(String)} method that converts the input
\code{String} to a token iterator (details omitted for brevity),
transitions according to that iterator, and then returns whether the
string is accepted. The goal is to synthesize an automaton that
recognizes \bcode{c(a$|$d)$+$r}.  The corresponding harness
\code{TestCADsR.examples()} in Figure~\ref{fig:sample} constructs a
\code{CADsR} instance and makes various assertions about its behavior.
Notice that this example relies critically on class generators, since
\code{Monitor} and \code{CADsR} must encode different automata.


\paragraph*{Output.}

\begin{figure}

\begin{lstlisting}[name=Ex]
class Automaton1 {
  int state = 0; static int num_state = 3;
  public void transition (Token t) { ...
    assert 0 <= state && state < 3;
    if (state == 0 && id == 1) { state = 1; return; } // open@
    if (state == 1 && id == 1) { state = 2; return; } // open 2x
    if (state == 1 && id == 2) { state = 0; return; } // (init)@
    if (state == 0 && id == 2) { state = 2; return; } // close 2x
  }
  public boolean accept() { return state <= 1; } ...
}
class DBConnection{ class Monitor extends Automaton1 { ... } ...}
class Automaton2 {
  int state = 0; static int num_state = 3;
  public void transition (Token t) { ...
    assert 0 <= state && state < 3;
    if (state == 0 && id == 99)  { state = 1; return; } // c
    if (state == 1 && id == 97)  { state = 2; return; } // ca
    if (state == 1 && id == 100) { state = 2; return; } // cd
    if (state == 2 && id == 114) { state = 0; return; } // c(a|d)+r@
  }
  public boolean accept() { return state <= 0; } ...
}
class CADsR extends Automaton2 { ... }
\end{lstlisting}

  \caption{\name{} Output (partial).}
  \label{fig:output}
\end{figure}

Figure~\ref{fig:output} shows the output produced by running \name{}
on the code in Figures~\ref{fig:template} and~\ref{fig:sample}.
We see that the generator was
instantiated as \code{Automaton1}, inherited by \code{DBConnection.Monitor},
and \code{Automaton2}, inherited by \code{CADsR}.
Both automata are equivalent to what we would expect for these
languages. Two things were critical for achieving this result:
minimizing the number of states (line~\ref{line:min-num-st}) and
having sufficient harnesses (Figure~\ref{fig:sample}).


We experimented further with \code{CADsR} to see how changing the
sketch and harness affects the output. First, we tried running with a
smaller harness, i.e., with fewer examples. In this case, the
synthesized automaton covers all the examples but not the full
language. For example, if we omit the four-letter inputs in
Figure~\ref{fig:sample} the resulting automaton only accepts
three-letter inputs. Whereas going to four-letter inputs constrains
the problem enough for \name{} to find the full solution.

Second, if we omit state minimization (line~\ref{line:min-num-st}),
then the synthesizer chooses large, widely separated indexes for
states, and it also includes redundant states (that could be merged
with a textbook state minimization algorithm).

Third, if we manually bound the number of states to be too small
(e.g., manually set \code{num\_state} to 2), the synthesizer runs for
more than half an hour and then fails, since there is no possible
solution.

Of these cases, the last two are relatively easy to deal with since
the failure is obvious, but the first one---knowing that a synthesis
problem is underconstrained---is an open research challenge. However,
one good feature of synthesis is that, if we do find cases that are
not handled by the current implementation, we can simply add those
cases and resynthesize rather than having to manually fix the code
(which could be quite difficult and/or introduce its own bugs).
Moreover, minimization---trying to ensure the output program is
small---seems to be a good heuristic to avoid overfitting to the
examples.

\section{Implementation}
\label{sec:java-sketch}


We implemented \name{} as a series of Python scripts
that invokes \sketch{} as a subroutine.
\name{} comprises roughly 5.7K lines of code,
excluding the parser and regression testing code.
\name{} parses input files
using the Python front-end of ANTLR v3.1.3~\cite{ANTLR} and its
standard Java grammar.
We extended that grammar to support holes, expression-level generators,
\code{minrepeat}, and the \code{harness} and \code{generator} modifiers.

There are a number of technical challenges in the implementation of
\name{}; due to space limitations we discuss only the major ones.

\paragraph*{Class hierarchy.}
The first issue we face is that
\sketch{}'s language is not object-oriented.
To solve this problem, \name{} 
follows a similar approach to~\cite{autograder} and
encodes objects with a new type
\code{V\_Object}, defined as a
\code{struct} containing all possible fields plus an
integer identifier for the class. More precisely, if $C_1, \ldots, C_m$ are
all classes in the program, then we define:
\begin{lstlisting}[name=Ex]
struct V_Object {
  int class_id; /**\it fields-from-$C_1$*/ ... /**\it fields-from-$C_m$*/
}
\end{lstlisting}
where each $C_i$ gets its own unique id.

\name{} also assigns every method a unique id, and 
it creates various constant arrays that record type information.
For a method id \code{m}, we set \code{belongsTo[m]} to be its class id;
\code{argNum[m]} to be its number of arguments; and
\code{argType[m][i]} to be the type of its \code{i}-th argument.  We
model the inheritance hierarchy using a two-dimensional array
\code{subcls} such that \code{subcls[i][j]} is true if class \code{i}
is a subclass of class \code{j}.

\paragraph*{Encoding names.}

When we translate the class hierarchy into \name{}, we also flatten
out the namespace, and we need to avoid conflating overridden or
overloaded method names, or inner classes.

Thus, we name inner classes as \code{\emph{Inner}\_\emph{Outer}},
where \code{\emph{Inner}} is the name of the nested class and
\code{\emph{Outer}} is the name of the enclosing class.  We also
handle anonymous classes by assigning them distinct numbers, e.g.,
\code{\emph{Cls}\_1}.

To support method overriding and overloading, methods are named
\code{\emph{Mtd}\_\emph{Cls}\_\emph{Params}},
where \code{\emph{Mtd}} is the name of the method,
\code{\emph{Cls}} is the name of the class in which it is declared, and
\code{\emph{Params}} is the list of parameter types.  For example,
in the finite automaton example, \code{CADsR} inherits method \code{transition} from \code{Automaton2} (the second variant of the class generator), hence the method is named
\code{transition\_Automaton2\_Token(V\_Object self, V\_Object t)} in \sketch{}.
The first parameter represents the callee of the method.

\paragraph*{Dynamic dispatch.} 

We simulate the dynamic dispatch mechanism of Java in \sketch{}.
For each method name \code{M} (suitably encoded, as above),
we introduce a function
\code{dyn\_dispatch\_M(V\_Object self, ...)} that dispatches based
on the \code{class\_id} field of the callee:
\begin{lstlisting}[name=Ex]
void dyn_dispatch_M(V_Object self, ...) {
  int cid = self.class_id;
  if (cid == R0_id) return M_R0_P(self, ...);
  if (cid == R1_id) return M_R1_P(self, ...);
  ...
  return;
}
\end{lstlisting}
Note that if \code{M} is static, the \code{self} argument is omitted.

\paragraph*{Java libraries.}

To perform synthesis, we need \sketch{} equivalents of any Java
standard libraries used in the input sketch.  Currently, \name{}
supports the following collections and APIs: \code{ArrayDeque},
\code{Iterator}, \code{LinkedList}, \code{List}, \code{Map},
\code{Queue}, \code{Stack}, \code{TreeMap}, \code{CharSequence},
\code{String}, \code{StringBuilder}, and \code{StringBuffer}. Library
classes are implemented using a combination of translation of the
original source using \name{} and manual coding, to handle native
methods or cases when efficiency is an issue.  Note that several of
these classes include generics (e.g., \code{List}), which is naturally
handled because the all objects are uniformly represented as
\code{V\_Object}.

\paragraph*{Limitations and unsupported features.}

As Java is a very large language, \name{} currently only supports
a core subset of Java. We leave several features of Java to the future versions
of \name{}, including packages, access control, exceptions, and concurrency.

Additionally, \name{} assumes the input sketch is type correct,
meaning the standard parts of the program are type correct, holes are
used either as integers or booleans, and expression generators are
type correct. This assumption is necessary because, although \sketch{}
itself includes static type checking, distinctions between different
object types are lost by collapsing them all into \code{V\_Object}.


\paragraph*{Using \sketch{}.}

We translate \name{} file, which is composed of the user-given template
and examples, as well as supportive libraries (if necessary) to \code{.sk}
files as input to \sketch{}.
For example,
the \code{SimpleMath} example in Section~\ref{sec:simplemath} is
translated to
\begin{lstlisting}[name=Ex]
int e_h1 = ??;
int mult2_SimpleMath_int(int x) { return e_h1 * {| x | 0 |}; }
harness void test_Test() { assert mult2_SimpleMath_int(3) == 6; }
\end{lstlisting}
We refer the reader elsewhere~\cite{sttt-sketch-paper, sketch:manual}
for details on how \sketch{} itself works.

 
After solving the synthesis problem, \name{}
then unparses these same Java files, but with unknowns
resolved according to the \sketch{} synthesis results.
We use partial parsing~\cite{Partial} to make this output process
simpler.

\section{Experience with \name{}}
\label{sec:experience}

We developed \name{} as part of the development of another tool,
\pasket{}~\cite{pasket}, 
which aims to construct \emph{framework models}, e.g. mock classes that
implement key functionality of a framework but in a way that is much simpler
than the actual framework code and is more amenable to static analysis.
\pasket{} takes as input a log of the interaction between the real framework
and a test application, together with a description of the API of the framework
and of the design patterns that the framework uses. \pasket{} uses these inputs
to automatically generate an input to \name{} which is then responsible for
actually synthesizing the models. Through \pasket{}, we have used \name{}
to synthesize models of key functionality from the Swing and Android frameworks.
The largest \name{} inputs generated by \pasket{} contain 117 classes and
4,372 lines of code, and solve in about two minutes despite having
over $73^{18} \times 164^{28}$ possible choices;
this is possible thanks to a new synthesis algorithm
called Adaptive Concretization~\cite{jeon:cav15}
that is available in \sketch{} and was also developed as part of this work.

\section*{Acknowledgments}

This research was supported in part by NSF CCF-1139021, CCF-1139056,
CCF-1161775, and the partnership between UMIACS and the Laboratory for
Telecommunication Sciences.

\bibliographystyle{abbrvnat}

\newpage

\appendix

\section{Tool Demonstration Walkthrough}
\label{sec:walkthrough}

As mentioned in the introduction, \name{} is available at
\url{http://plum-umd.github.io/java-sketch/}.  The tool is at a
fairly early stage of development, but is robust enough to be used by
the wider research community.

Our demonstration will generally follow the overview in
Section~\ref{sec:overview}. Below are more details of what we plan to
present.

\subsection{Basics}

\name{} performs \emph{program synthesis}---generating an output
program given an input specification.

Let's begin with a small example:
\begin{small}
\begin{verbatim}
$ cat >> SimpleMath.java
class SimpleMath {
    static int mult2(int x) {
        return ?? * {| x , 0 |};
    }
}
\end{verbatim}
\end{small}
\noindent
This is a \emph{sketch} (also \emph{scaffold} or \emph{template}),
which is a \emph{partial Java program}. The \code{??} is a
\emph{hole}---unknown integer---and the other part of the product is a
\emph{generator}---ranging over the listed expressions. Notice that
this sketch has $2^{33}$ possible instantiations.

In addition to the template, the other important input to \name{} is
\emph{examples} that specify the expected behavior of the
template. These are analogous to unit tests.  We provide a
\emph{harness} containing \code{assert}ions about the \code{mult2}
method:
\begin{small}
\begin{verbatim}
$ cat >> Test.java
class Test {
    harness static void test() {
        assert SimpleMath.mult2(3) == 6;
    }
}
\end{verbatim}
\end{small}
Now we can run \name{} on the sketch and harness:
\begin{small}
\begin{verbatim}
$ ./jsk.sh SimpleMath.java Test.java
06:07:15 rewriting syntax sugar
06:07:15 specializing class-level generator
06:07:15 rewriting exp hole
06:07:15 semantics checking
06:07:15 building class hierarchy
06:07:15 encoding result/sk_Test/SimpleMath.sk
06:07:15 encoding result/sk_Test/Test.sk
...
06:07:15 sketch running...
06:07:15 sketch done: result/output/Test.txt
06:07:15 replacing holes
06:07:15 replacing generators
06:07:15 semantics checking
06:07:15 decoding result/java/SimpleMath.java
...
06:07:15 synthesis done
\end{verbatim}
\end{small}

The final result synthesized by \name{} is a valid Java source file
where unknowns have been replaced with the appropriate code:
\begin{small}
\begin{verbatim}
$ cat result/java/SimpleMath.java
class SimpleMath { ...
static int mult2 (int x) {
return 2 * x;
}
}
\end{verbatim}
\end{small}

\subsection{Database Connection Monitor}

Now consider a harder problem: suppose we want to synthesize an automaton-based
inline reference monitor to check basic properties of a database
connection, namely that the connection is never opened or closed twice
in a row and is only closed after being opened.
Let's use a simple, efficient implementation: representing states via
distinct integers, along with a series of conditionals that encode state
transitions.

Here's the initial sketch:
\begin{lstlisting}[name=Ex2]
interface Token { public int getId(); } /** \label{line2:token} */
class Automaton {
    private int state = ??; /** \label{line2:init-st} */
    static int num_state = ??; /** \label{line2:num-st} */
    harness static void min_num_state() { minimize(num_state); } /** \label{line2:min-num-st} */
    public void transition(Token t) {
        assert 0 <= state && state < num_state; /** \label{line2:chk-st} */
        int id = t.getId(); /** \label{line2:getid} */
        minrepeat { /** \label{line2:minrepeat} */
            if (state == ?? && id == ??) { state = ??; return; } /** \label{line2:transition} */
    }   }
    public void transitions(Iterator<Token> it) {
        while(it.hasNext()) { transition(it.next()); }
    }
    public boolean accept() { return state <= ??; } /** \label{line2:accept} */
}
\end{lstlisting}
\noindent
Here are some key things to notice about the source code:
\begin{itemize}
\item transitions are taken based on the \code{Token} interface
\item the initial state is arbitrary (line~\ref{line2:init-st})
\item the number of states is arbitrary (line~\ref{line2:num-st})
\item states are dense (line~\ref{line2:min-num-st})
\item we use an assertion to check the validity of the current state
  (line~\ref{line2:chk-st})
\item a transition is arbitrary as it depends on an arbitrary current state
  and an arbitrary id (line~\ref{line2:transition})
\item \code{minrepeat} replicates its body the minimum necessary
  number of times.
\item the number of transitions is arbitrary (line~\ref{line2:minrepeat})
\item the number of accepting states is arbitrary
  (line~\ref{line2:accept}); by packing the states densely, we could
  use an inequality here
\end{itemize}

Now we can define an inline reference monitor as follows:
\begin{lstlisting}[name=Ex2]
class DBConnection {
    class Monitor extends Automaton {
        final static Token OPEN =
            new Token() { public int getId() { return 1; } };
        final static Token CLOSE =
            new Token() { public int getId() { return 2; } };
        public Monitor() { }
    }
    Monitor m;
    public DBConnection() { m = new Monitor(); }
    public boolean isErroneous() { return ! m.accept(); }
    public void open() { m.transition(Monitor.OPEN); }
    public void close() { m.transition(Monitor.CLOSE); }
}
\end{lstlisting}
\noindent
The key idea is that each database connection operation is
associated with an unique id, and the monitor maintains an
automaton that keeps receiving operation ids.
At any point, a client can check the status of the connection
by asking the monitor whether it is in an accepting state.

As expected, we need to provide a harness:
\begin{lstlisting}[name=Ex2]
class TestDBConnection {
    harness static void scenario_good() { /** \label{line:harness-good} */
        DBConnection conn = new DBConnection();
        assert ! conn.isErroneous();
        conn.open();  assert ! conn.isErroneous();
        conn.close(); assert ! conn.isErroneous();
    }
    // bad: opening more than once
    harness static void scenario_bad1() { /** \label{line:harness-bad1} */
        DBConnection conn = new DBConnection();
        conn.open(); conn.open(); assert conn.isErroneous();
}   }
\end{lstlisting}
\noindent
These examples illustrate one normal usage---opening a connection
and closing it---and one abnormal usage---opening a connection twice.
Given these harnesses, \name{} finds a solution:
\begin{lstlisting}[name=Ex2]
class Automaton {
  int state = 0;
  static int num_state = 3;
  public void transition (Token t) { ...
    assert 0 <= state && state < 3;
    if (state == 0 && id == 1) { state = 1; return; } // open@
    if (state == 1 && id == 1) { state = 2; return; } // open 2x
  }
  public boolean accept() { return state <= 1; } ... }
\end{lstlisting}
\noindent

This sort of looks okay, but it's odd that there are no transitions
for the \code{close} operation.  When the monitor is in state 1
and the given operation is \code{close},
it is fine for the monitor to stay at the same accepting state.
But, it is problematic if we close the connection more than once,
which we should have specified:

\begin{lstlisting}[name=Ex2]
class TestDBConnection { ...
    // bad: closing more than once
    harness static void scenario_bad2() { /** \label{line:harness-bad2} */
        DBConnection conn = new DBConnection();
        conn.open();
        conn.close(); conn.close(); assert conn.isErroneous();
}   }
\end{lstlisting}

After adding that abnormal case---closing twice,
\name{} finds this solution:
\begin{lstlisting}[name=Ex2]
class Automaton {
  int state = 0;
  static int num_state = 3;
  public void transition (Token t) { ...
    assert 0 <= state && state < 3;
    if (state == 0 && id == 1) { state = 1; return; } // open@
    if (state == 1 && id == 1) { state = 2; return; } // open 2x
    if (state == 1 && id == 2) { state = 0; return; } // (init)@
    if (state == 0 && id == 2) { state = 2; return; } // close 2x
  }
  public boolean accept() { return state <= 1; } ... }
\end{lstlisting}
\noindent
The resulting automaton is exactly same as what one can write by hand.

\subsection{A Regular Language: Lisp-Style Identifier}

Now let's consider synthesizing another automaton, trying to create a finite
automaton given sample accepting and rejecting inputs.
One benefit of Java as an object oriented language is
code reuse via subclassing, so we could just make another class that
extends \code{Automaton}, assuming we want this to be part of the same
program. But subclassing won't quite work here because we need
different automata for each use case.

To solve this problem, we can make \code{Automaton} a \emph{class generator},
so that it can be instantiated differently in different
superclass contexts.
\begin{lstlisting}[name=Ex2]
generator class Automaton { ... }
class DBConnection {
    class Monitor extends Automaton { ... }
    ...
}
class CADsR extends Automaton { ... }
\end{lstlisting}

Now let's use this to synthesize an example automaton:
\begin{lstlisting}[name=Ex2]
class CADsR extends Automaton { ...
    public boolean accept(String str) {
        state = init_state_backup;
        transitions(convertToIterator(str));
        return accept();
}   }
\end{lstlisting}
\noindent
Note the overloaded \code{accept(String)} method.
Now, we need to specify sample strings that are accepted or rejected
by the synthesized automaton.  Suppose we want to synthesize
an automaton that recognizes Lisp-style identifier \bcode{c(a$|$d)$+$r}.
The following harness constructs a \code{CADsR} instance and makes
several assertions about its behavior:
\begin{lstlisting}[name=Ex2]
class TestCADsR {
    harness static void examples() {
        CADsR a = new CADsR();
        assert ! a.accept("c");  assert ! a.accept("cr");
        assert a.accept("car");  assert a.accept("cdr");
        assert a.accept("caar"); assert a.accept("cadr");
        assert a.accept("cdar"); assert a.accept("cddr");
}   }
\end{lstlisting}
\noindent
If we provide less examples, e.g., if we remove examples
about rejected strings, the synthesizer simply returns an automaton
that does not make any transitions, while the initial state is an
accepting state.  This awkward automaton actually conforms to
any accepted strings, and one can easily figure out the necessity
of rejected strings.

To see the advantage of using \code{minimize},
let's run synthesis without line~\ref{line2:min-num-st}.
We get:
\begin{lstlisting}[name=Ex2]
class Automaton2 {
  int state = 106; static int num_state = 120;
  public void transition (Token t) { ...
    if (state == 106 && id == 99) { state = 64; return; }  // c
    if (state == 64 && id == 97) { state = 100; return; }  // ca
    if (state == 100 && id == 114) { state = 50; return; } // car@
    if (state == 64 && id == 100) { state = 119; return; } // cd
    if (state == 119 && id == 114) { state = 32; return; } // cdr@
    if (state == 64 && id == 114) { state = 72; return; }  // cr
  }
  public boolean accept() { return state <= 50; } ... }
\end{lstlisting}
\noindent
Notice that the synthesizer picked fairly strange numbers for the
states and left a lot of states unused. Moreover, the automaton is
inefficient in that it uses two different paths and final states to
accept ``\code{car}'' and ``\code{cdr}''.

If we run synthesis again using \code{minimize}, we get:
\begin{lstlisting}[name=Ex2]
class Automaton2 {
  int state = 0; static int num_state = 3;
  public void transition (Token t) { ...
    assert 0 <= state && state < 3;
    if (state == 0 && id == 99)  { state = 1; return; } // c
    if (state == 1 && id == 97)  { state = 2; return; } // ca
    if (state == 1 && id == 100) { state = 2; return; } // cd
    if (state == 2 && id == 114) { state = 0; return; } // c(a|d)+r@
  }
  public boolean accept() { return state <= 0; } ... }
\end{lstlisting}
\noindent
This result is better in the sense that it uses dense states and
that it encompasses only one path and final state to accept the valid strings.


To double-check whether it is indeed the minimum number of states,
we can test with the bounded number of states:
\begin{lstlisting}[name=Ex2]
class Automaton {
    static int num_state = 2; ...
}
\end{lstlisting}
\noindent
In this case, the synthesizer runs for more than half an hour and
then fails, as there is no possible solution using only two states.



\subsection{Internals of \name{}}

If time permits, we will show a bit of \name{}'s translation to
\sketch{}. For example, the translation of the \code{mult2} example
looks like:
\begin{small}
\begin{verbatim}
$ cat result/sk_Test/SimpleMath.sk
...
int e_h1 = ??;
int mult2_SimpleMath_int(int x) {
return e_h1 * {| x | 0 |};
}
$ cat result/sk_Test/Test.sk
...
harness void test_Test() {
assert mult2_SimpleMath_int(3) == 6;
}
\end{verbatim}
\end{small}

\name{} extracts the synthesis results by looking at how each hole was
solved by \sketch{}:
\begin{small}
\begin{verbatim}
$ cat result/log/log.txt
...
06:07:15 [DEBUG] java_sk/decode/finder.py:41
hole: SimpleMath.e_h1
06:07:15 [INFO] java_sk/decode/__init__.py:69
replacing holes
06:07:15 [DEBUG] java_sk/decode/replacer.py:72
replaced: SimpleMath.e_h1 = 2
06:07:15 [DEBUG] java_sk/decode/replacer.py:89
replaced: e_h1 @ int SimpleMath.mult2(int) with 2
06:07:15 [DEBUG] java_sk/decode/finder.py:93
generator@mult2: {| x | 0 |}
06:07:15 [INFO] java_sk/decode/__init__.py:79
replacing generators
06:07:15 [DEBUG] java_sk/decode/replacer.py:151
{| x | 0 |} => x
...
\end{verbatim}
\end{small}

Then it traverses the original Java sketch and outputs it, plugging in
the solved values for the holes.

\end{document}